\documentclass[%
 reprint,
superscriptaddress,
 amsmath,amssymb,
prx,
]{revtex4-2}

\usepackage{graphicx}% Include figure files
\usepackage{dcolumn}% Align table columns on decimal point
\usepackage{bm}% bold math
\usepackage[colorlinks]{hyperref}

\begin{document}

\preprint{APS/123-QED}

\title{Quantum-Referenced Spontaneous Emission Tomography}

\author{I. I. Faruque}
\email{imad.faruque@bristol.ac.uk}
\affiliation{Quantum Engineering Technology Laboratories, University of Bristol, Bristol, United Kingdom}
\author{B. M. Burridge}
\affiliation{Quantum Engineering Technology Laboratories, University of Bristol, Bristol, United Kingdom}
\affiliation{Quantum Engineering Centre for Doctoral Training, Centre for Nanoscience \& Quantum Information, University of Bristol, Bristol, United Kingdom}
\author{M. Banic}
\affiliation{Department of Physics, University of Toronto, 60 St. George Street, Toronto, Ontario, Canada, M5S 1A7}
\author{M. Borghi}
\affiliation{Dipartimento di Fisica, Università di Pavia, via Bassi 6, 27100, Pavia, Italy}
\author{J. E. Sipe}
\affiliation{Department of Physics, University of Toronto, 60 St. George Street, Toronto, Ontario, Canada, M5S 1A7}
\author{J. G. Rarity}
\affiliation{Quantum Engineering Technology Laboratories, University of Bristol, Bristol, United Kingdom}
\author{J. Barreto}
\affiliation{Quantum Engineering Technology Laboratories, University of Bristol, Bristol, United Kingdom}
%\affil[*] {imad.faruque@bristol.ac.uk , $^\dagger$ These authors contributed equally}

\date{\today}% It is always \today, today,

\begin{abstract}
We present a method of tomography that measures the joint spectral phase (JSP) of spontaneously emitted photon pairs originating from a largely uncharacterized ``target" source. We use quantum interference between our target source and a reference source to extract the JSP with four spectrally resolved measurements, in a process that we call quantum-referenced spontaneous emission tomography (Q-SpET). We have demonstrated this method on a photonic integrated circuit for a target micro-ring resonator photon-pair source. Our results show that spontaneously emitted photon pairs from a micro-ring resonator are distinctively different from that of stimulated emission, and thus cannot in general be fully characterized using classical stimulated emission tomography without detailed knowledge of the source. %SpET does not require exotic optical components, complex filtering techniques or extensive processing algorithms. 

%By setting up a quantum interferometer between our target and a reference source, we extract the joint spectral phase using as little as four measurements. Our method; which we call spontaneous emission tomography (SpET), is entirely self-referential, requires no exotic components or filtering techniques and calls for negligible processing. We have demonstrated this method on a photonic integrated circuit for a micro-ring resonator photon-pair source, ensuring inherent phase stability. We have experimentally shown for the first time that spontaneously emitted photon-pairs from a micro-resonator are distinctively different from that of stimulated emission, and thus cannot be fully characterized using classical stimulated emission tomography.
\end{abstract}

\maketitle
%Notes so far:
%   Change tone, present what we have done.
%   JSP + No analytical model + Self-referencing + Low complexity + Differences from SET.#

%Re-written Intro + Abstract.

\section{Introduction}
% What sources require models that we cannot have? - topological sources
% What applications require full characterisation of photon-pair sources
% What other application may be benefited from a phase tomography method 
%Para1: Photon-pair sources are essential in time-frequency entanglement, nonlocal quantum interference phenomena, quantum computing, quantum simulator, quantum key distribution, quantum metrology, sensing and nonlinear spectroscopy. 
%Para2: The generation and control of photon-pairs has been realized in bulk crystals, optical fibres and chip-scale waveguides and micro-resonators; however, it remains an important challenge to reconstruct the quantum state of the photons produced for all the different sources.
Single photons are one of the simplest and more powerful quantum information carriers harnessed by quantum technologies. As flying qubits, single photons are the workhorse for applications in quantum communications such as global, entanglement-based quantum networks \cite{Ekert91, Micius_entanglement, SKJoshi2018, Multi_dimension, RevHom2}, while also enabling direct and indirect interactions with the environment for metrology, sensing and imaging applications \cite{Metrology_review, Lemos2014}. Moreover, integrated photonic circuits using single photons are one of the leading candidates for performing quantum information processing tasks \cite{LOQC, Rudolph2017, Gentile2021}. 
%An effective method of obtaining single photons is by heralding a photon generation event that instead creates a pair of entangled photons.
%furthermore quantum information processing techniques may make use of single photons, and entangled photons to implement protocols that go beyond classical information techniques. Therefore entangled states of single photons are a valuable resource for many quantum technologies.
Non-linear, parametric processes are at the heart of all-optical photon pair generation, and are a promising means to create heralded single photons \cite{pittman2005heralding}. Heralding mitigates the vacuum contributions associated with the probabilistic generation of photons, enables entanglement distribution protocols \cite{Micius_entanglement}, and produces what is effectively a single photon source. However, as a parametric process, heralding in the presence of correlation between the photons of a pair can project the remaining photon into a mixed state. Therefore the accurate characterization of photon-pair sources is essential. 

% a prerequisite in instances such as the use of photon entanglement to enhance metrology beyond its classical limits \cite{Metrology_review, Lemos2014}, and for the development of global, entanglement-based quantum networks \cite{Ekert91, Micius_entanglement, Satellite_QKD_review, Multi_dimension, RevHom2}.
%3) Chip-scale quantum photonics - topological photon-pairs etc.
%(3rd para is how/what we have done)
%(4th para is the lit rev)

Photon pairs can be described by their biphoton wave function, or joint spectral amplitude (JSA), a quantity that depends on the frequencies of the two photons. The square magnitude of the JSA --- the joint spectral intensity (JSI), can be measured directly using single-photons and a scanning monochromator \cite{Kim2005} or more rapidly using stimulated emission tomography (SET) that makes use of commercial optical characterization equipment \cite{Liscidini2013, Grassani2016}. However, the JSA is a complex-valued function, requiring both amplitude and phase information for a complete description. Typically, the assumption is made that the JSI can be used to infer the spectral correlations of the JSA \cite{Grassani2016}, resulting in an upper bound on spectral correlation or entanglement. The argument of the JSA --- the joint spectral phase (JSP) describes the phase correlation of photon pairs, which also contributes to their entanglement. Therefore, the full JSA is required to fully characterize the entanglement between photon pairs of a specific source. Moreover, it has recently been shown that the JSP of photon pairs generated from resonant sources cannot be extracted by using SET alone \cite{St_JSP}. If an analytical model of photon-pair generation inside the resonant source exists, it may be combined with SET measurements to provide an estimate of the correct JSP. This may not be feasible for a complex resonant source such as topological sources \cite{Mittal2021}. 
%A common thread among methods that offer SET to overcome photon-counting limitations.
%Mtehods like Thekkadeth use three-fold coincidences to retrieve the JSP, which is a time-consuming process. They propose SET as a means to massively speed up the measurement, but this fails to identify, or address concerns raised in [15].

%Despite attempts to overcome the limitations of SET such as the use of time-consuming three-fold coincidences [Thekkadeth] the concerns raised in [15] have not yet been fully addressed.

%%%%%%%%%%%%%%%%%%%%%%%%%%%%%%%%%%%%%%%%%%%%%%%%%%
\begin{figure*}[t]
\centering
\includegraphics[width=\textwidth]{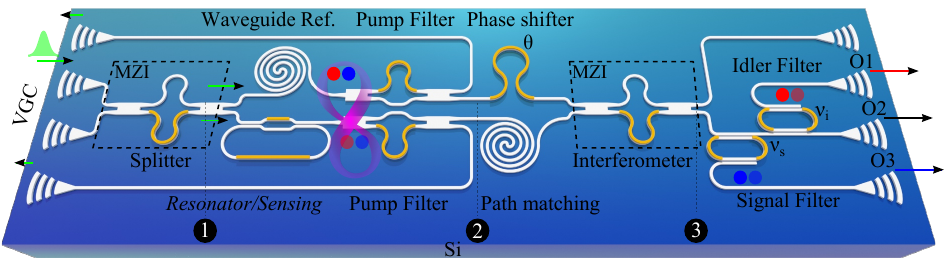}
\caption{Schematic of our photonic chip. VGC structures are used as optical input and output (O1, O2, O3) ports. The circled numbers (1, 2, 3) refer to the kets ($|\Psi\rangle$) at specific places in the photonic circuit. The optical pump is distributed to the spiral and the resonator element using an MZI ($|\Psi_1\rangle$). On-chip AMZI filters are then used to remove the pump from the optical channel ($|\Psi_2\rangle$). A phase $\theta$ can then be applied to one of the arms using a thermal phase-shifter, before the optical path compensation (i.e., matching) element. The photon states generated at the spiral and resonator are brought together and interfered using an MZI ($|\Psi_3\rangle$). Two additional tuneable on-chip optical filters based on resonator structures allow for performing frequency-resolved measurements.}
\label{fig:chip}
\end{figure*}
%%%%%%%%%%%%%%%%%%%%%%%%%%%%%%%%%%%%%%%%%%%%%%%%

There have been some attempts to remedy this separately, with methods previously reported for both single photons, and photon pairs \cite{davis2018measuring, Davis2020} using Electro-Optic Shearing Interferometry (EOSI), which enforces tight timing synchronization, and phase-stabilization \cite{beduini2014interferometric}. Experimental complexity can be substituted for large processing overheads associated with artificially constrained phase-retrieval algorithms \cite{MacLean2019}. Alternatively, the methodology can be simplified under the assumption that the photons are highly correlated \cite{Tischler2015}. Methods that use SET to overcome photon-counting limitations \cite{Thekkadath2022}, do not identify or address concerns raised in \cite{St_JSP}. Nevertheless, it is possible to extract the correct JSP using a self-referenced method \cite{Davis2020} by implementing Electro-Optic Shearing Interferometry as well as additional post-processing using Fourier filtering.

% (... to characterise one another)
Here we investigate a technique that recovers the JSP through the use of quantum interference between our target source, or ``Device-Under-Test" (DUT), and a reference source. Our DUT is a micro-resonator and our reference is a waveguide source. A salient and necessary feature of our method is the utilization of two dissimilar sources, as we rely on the resulting interference pattern to characterize our target device; in direct contrast to most experiments reported in the literature that have been performed with photons generated by identical sources (e.g., \cite{HOM, Multi_dimension, Wang2018,Ekert1992,NOON_theory, Fukuda:05,Jeffrey2004,Migdall2002,Faruque2018}). 

%to use?? \cite{MacLean2018} % JTA
Our technique is implemented entirely on-chip - showing the useability of our method for chip-scale quantum photonic technologies. We use a micro-racetrack resonator as a ``proof of principle" because it is a conventional and well-understood optical device and one of the most versatile, bright, and efficient photon-pair sources in integrated quantum photonic technologies \cite{LLewellyn2020, zhang2021squeezed, Micius_entanglement}.
For our reference, we use a waveguide photon-pair source to eliminate the need for a phase-stable, classical reference. The quantum nature of our reference ensures inherent spectral overlap with our target, without the need to enforce it with a complex combination of an external coherent state, and phase-stabilization \cite{Phase_tomography,beduini2014interferometric}.

Q-SpET's constraints are that the DUT spectra must overlap well with that of our reference, and any feature (such as chirp) in the profile of the pump should not vary significantly across the spectrum of the DUT. This ensures that the waveguide JSA has a flat phase profile and can be a suitable reference for the DUT.

\section{Implementing Spontaneous Emission Tomography}%SpET Tomography Using a Photonic Integrated Circuit
In our photonic chip, as can be seen in Fig.~\ref{fig:chip},  the pump light (described by a coherent state $|\alpha\rangle$) enters from the left through a vertical grating coupler (VGC), and its power is then divided across two paths according to the intensity ratio $\eta$ set by a Mach-Zehnder Interferometer (MZI). The pump light in each channel can lead to the production of photon-pairs through spontaneous four-wave mixing (SFWM) in the structure that follows. 
%The path leads, as shown in Fig.~\ref{fig:chip}, for the 
The upper channel leads to a spiral waveguide (W), and %for 
the lower channel to a micro-ring resonator (R). The quantum state exiting the MZI is represented by the circled number 1 in Fig.~\ref{fig:chip}, and the associated ket can be expressed as
\begin{align}
|\Psi_1\rangle = |\sqrt{\eta}\alpha\rangle_W|i\sqrt{1-\eta}\alpha\rangle_R.
\label{eq:pump_split}
\end{align}
If we use normalized biphoton wavefunctions $\Phi_R(\nu_s, \nu_i)$ and $\Phi_W(\nu_s, \nu_i)$ to characterize the probability amplitudes for the generation of pairs in the resonator and waveguide respectively, where $\nu_s$ and $\nu_i$ are signal and idler frequencies, then the normalized kets associated with these generation processes are \cite{Bi_photon}
\begin{align}
|\psi_{II}\rangle_{k} = \int d\nu_s d\nu_i \Phi_k(\nu_s,\nu_i ) a^{\dagger}_s(\nu_s) a^{\dagger}_i(\nu_i) |\mathrm{vac}\rangle, \label{eq:bi_photon_1}
\end{align}
where $k=\{R,W\}$, $\nu_s$ and $\nu_i$ are signal and idler frequencies, $a^{\dagger}_s(\nu_s)$ and $a^{\dagger}_i(\nu_i)$ are creation operators for signal and idler photons, and $|\mathrm{vac}\rangle$ is the vacuum state. 

%%%%%%%%%%%%%%%%%%%%%%%%%%%%%%%%%%%%%%%%%%%%%%%%%%
\begin{figure*}[t]
\centering
\includegraphics[width=\textwidth]{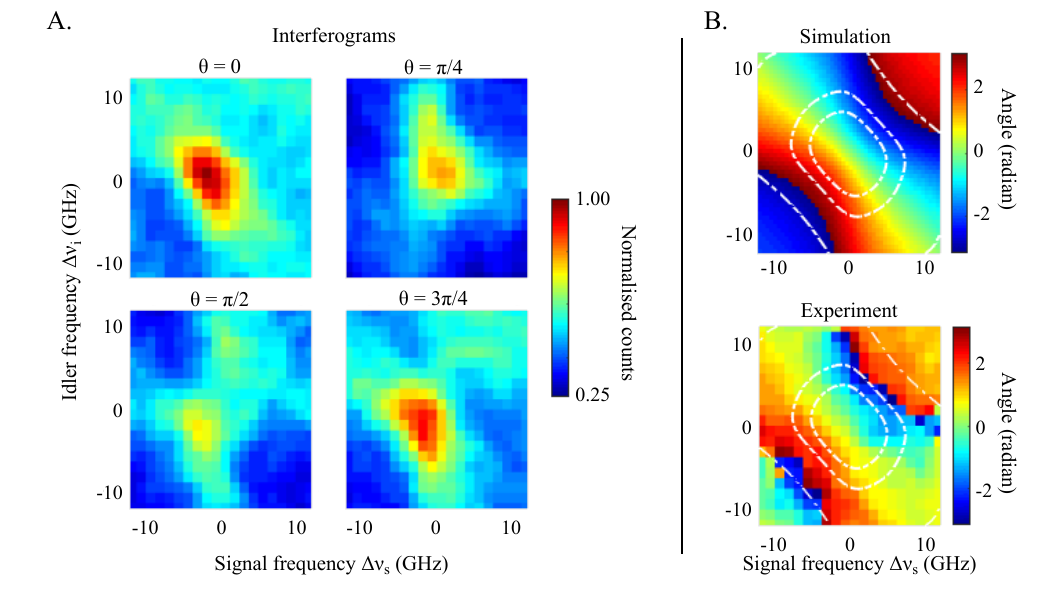}
\caption{Spontaneous emission tomography. \textbf{A}. Four measured interferograms at $\theta = \{0,\pi/4,\pi/2,3\pi/4\}$. A low-pass filter is used here to remove high-frequency noise.
\textbf{B}. Simulated and experimental JSP for spontaneously emitted photon pairs generated from a micro-resonator. %The experimental JSP is constructed from the measured interferograms using Eq.~\ref{eq:JSP}.} 
The superimposed white dotted contours represent constant intensity lines of the JSI (Fig.~\ref{fig:StJSP}{B}), indicating 25\%, 10\%, and 1\% of the maximum signal.}
\label{fig:SpJSP}
\end{figure*}
%%%%%%%%%%%%%%%%%%%%%%%%%%%%%%%%%%%%%%%%%%%%%%%%
An important condition for implementing Q-SpET is a restriction to the weak pump regime \cite{RevHom1, RevHom2}, where the pump power is low enough that the probability of more than one photon pair generation event is negligible. Then for each photon pair, there is a probability amplitude that it is generated in the waveguide and a probability amplitude that it is generated in the resonator. After the photon pair generation process, we use asymmetric MZIs to filter out the pump, resulting in a quantum state (circled number 2 in Fig.~\ref{fig:chip}) consisting mainly of the vacuum state with a small amount of the two-photon state $|\Psi_{2}\rangle$
\begin{align}
|\Psi_2\rangle = N_R(\eta)|\psi_{II}\rangle_{R}|\mathrm{vac}\rangle_{W} + N_W(\eta)|\mathrm{vac}\rangle_{R}|\psi_{II}\rangle_{W},
\label{eq:joint_state2a}
\end{align}
where the factors $N_R$ and $N_W$ depend on the intensity ratio $\eta$, and $|N_R|^2 + |N_W|^2 = 1$. The photon pairs from the resonator then propagate through a path matching waveguide section, and a relative phase ($\theta$) is applied using a thermo-optic phase-shifter to this arm of the circuit with respect to the waveguide arm.

Next, both arms of the circuit are connected to an MZI configured as a balanced beam-splitter. The result (See Supplemental Material) is a state which is sensitive to the interference between the two biphoton wavefunctions.

After this MZI, we use two on-chip, resonator-based tuneable optical filters to select the frequencies of light that we divert towards a given output. Photons filtered towards O3 and O1 are labeled signal and idler, respectively, and the remaining photons propagate into O2. All three outputs are finally coupled out of the chip using VGCs, and signal and idler photons are detected using off-chip superconducting single-photon detectors (Fig.~\ref{fig:setup}). The probability of detecting both a signal and idler of specific frequency in arms O3 and O1, given a phase angle $\theta$ and an intensity ratio $\eta$, involves a superposition of the amplitudes $\Phi_R(\nu_s, \nu_i)$, and $\Phi_W(\nu_s, \nu_i)$ and can be expressed as:

\begin{widetext}

\begin{align}
P\left( \theta, \eta, \nu_s, \nu_i \right) &= \frac{1}{4} \left| N_R(\eta)\Phi_R(\nu_s, \nu_i) e^{i2\theta} +  N_W(\eta)\Phi_W(\nu_s, \nu_i) \right|^2. \label{eq:probability}\\
\frac{P(\theta=\frac{3\pi}{4}) - P(\theta=\frac{\pi}{4})}{P(\theta=0) - P(\theta=\frac{\pi}{2})} &= \frac{N_RN_W|\Phi_W| \mathrm{Im}\{\Phi_R\}}{N_RN_W|\Phi_W| \mathrm{Re}\{\Phi_R\}} = \frac{\mathrm{Im}\{\Phi_R\}}{\mathrm{Re}\{\Phi_R\}}= \tan(\theta_{\mathrm{JSA}}). \label{eq:JSP}
\end{align}

\end{widetext}

Our method of tomography uses a specific set of $\theta$ values ($\{0,\pi/4,3\pi/4,\pi/2\}$). At each setting of $\theta$ we sweep the on-chip filters over a range of frequencies ($\nu_s$ and $\nu_i$) and measure the coincidences, which, once normalized, can be directly related to the probability (Eq.~\ref{eq:probability}). This process results in the four interferograms seen in Fig.~\ref{fig:SpJSP}{A}. The resolution of our data is governed by the bandwidth of our on-chip filters (10~pm), affording us a high level of precision in the single photon regime
. We can use simple algebraic manipulation (Eq.~\ref{eq:JSP}) on the measured data to recover the JSP (denoted by $\theta_{\mathrm{JSA}}$) as expressed in Eq.~\ref{eq:JSP}.

One significant feature of Q-SpET is that the estimated JSP does not depend on the normalization factors ($N_R$, $N_W$), and thus the method is independent of the intensity ratio $\eta$. This can be seen in Eq.~\ref{eq:JSP}, as all the pre-factors cancel out, leaving only the real ($\mathrm{Re}\{\Phi_R\}$) and imaginary ($\mathrm{Im}\{\Phi_R\}$) parts of the JSA.

%\newpage
%\begin{strip}
%\begin{align}
%\frac{P(\theta=\frac{3\pi}{4}) - P(\theta=\frac{\pi}{4})}{P(\theta=0) - P(\theta=\frac{\pi}{2})} = \frac{N_RN_W|\Phi_W| \mathrm{Im}\{\Phi_R\}}{N_RN_W|\Phi_W| \mathrm{Re}\{\Phi_R\}} = \frac{\mathrm{Im}\{\Phi_R\}}{\mathrm{Re}\{\Phi_R\}}= \tan(\theta_{\mathrm{JSA}}). \label{eq:JSP}
%\end{align}
%\end{strip}

%%%%%%%%%%%%%%%%%%%%%%%%%%%%%%%%%%%%%%%%%%%%%%%%%%
\begin{figure*}[t]
\centering
\includegraphics[width=\textwidth]{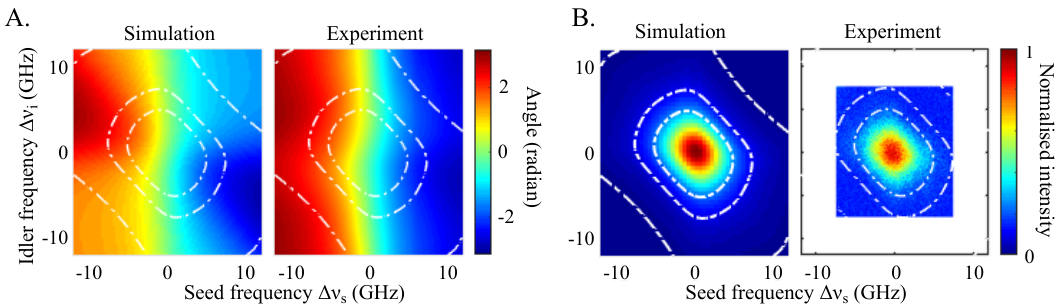}
\caption{Stimulated emission tomography. Simulated and experimental \textbf{A} - JSP and \textbf{B} - JSI of the StFWM in a micro-resonator. The white contour lines represent the 25\%, 10\%, and 1\% intensity bounds as in Fig.~\ref{fig:SpJSP}{B}. The experimental signal is primarily detectable within the innermost contour. Considering very weak signal is present outside this contour, the experimental JSPs have remarkable similarity within the outer contours. Data was not collected in the white region in the JSI plot.}
\label{fig:StJSP}
\end{figure*}
%%%%%%%%%%%%%%%%%%%%%%%%%%%%%%%%%%%%%%%%%%%%%%%%%%

Fig~\ref{fig:SpJSP}{B} shows both a simulated JSP obtained following the method described %in 
earlier \cite{two_strategies} and the JSP that we have obtained from experimental measurements (Fig~\ref{fig:SpJSP}{A}). The JSPs in Fig~\ref{fig:SpJSP}{B} additionally include white `confidence' contours indicating 25\%, 10\%, and 1\% of the maximum signal. We find reasonably good agreement between simulated and measured data, especially within the two innermost contours (signal above 10\% of the maximum). Over the frequency range observed for the resonator JSA (25~GHz), the waveguide JSA varies so little that it can indeed serve as a good reference \cite{Faruque2019} (See Supplemental Material).

Very generally, once the JSP is determined in this way, and the JSI identified by direct detection of the frequency dependence of the emitted signal and idler photons, the full JSA can be constructed.

%, in addition to the spontaneous measurements of the JSP (Fig.~\ref{fig:SpJSP}{B}),
%We then use the SET data to obtain stimulated measurements of the JSP (see Fig.~\ref{fig:StJSP}{A}). To obtain the Stimulated FWM (StFWM) interferograms experimentally we sweep the wavelength of an external seed laser and record the spectra of the stimulated idler photons at O2 (see Fig.~\ref{fig:chip}).} Then, by selecting specific values of $\theta$, we are able to estimate the JSP of the StFWM process.
%%%%%%%%%%%%%%%%%%%%%%%%%%%%%%%%%%%%%%%%%%%%%%%%

\section{Stimulated Vs Spontaneous Emission Tomography of a resonant process} % SpET vs SET

Of course, SET can be used to determine the JSI much more easily than direct detection of the spontaneous process \cite{Liscidini2013, St_JSP}. We have done this for our integrated circuit (Figs.~\ref{fig:setup}, \ref{fig:StJSP_interferograms}) using a seed at signal frequencies and recording the stimulated intensity at the idler frequencies. The results are shown in Fig.~\ref{fig:StJSP}{B} together with a simulation; again there is reasonably good agreement between the measured and simulated data. The resolution of our SET data, 125~MHz (1~pm) and 20~MHz (0.16~pm) for the seed and idler frequencies respectively, is at least an order of magnitude better than the existing literature \cite{Grassani2016}.

Interestingly, just as interferograms were constructed in the \textcolor{red}{Q-}SpET experiment, we can construct stimulated FWM (StFWM) interferograms, here by sweeping the wavelength of an external seed laser and recording the spectra of the stimulated idler photons at O2 (see Fig.~\ref{fig:chip}). %We then, by 
By using the interferograms at specific values of $\theta$, we can then extract a JSP of the StFWM process, (See Supplemental Material) analogously to what was done for the spontaneous process (Fig~\ref{fig:SpJSP}{B}). And again, there is a reasonably good agreement between the measured data and the simulation (Fig~\ref{fig:StJSP}{A}). 

However, while the JSI determined by the spontaneous and stimulated processes should agree -- that, of course, being the basis of SET -- the JSP obtained from the two processes are in general different. This illustrates how a measured JSI can deviate from that of simulation (Fig.~\ref{fig:SpJSP}a, Supplemental Material, \cite{Rozema2015}).%his allows us to compare directly the signatures of the stimulated, and spontaneous four-wave mixing processes. For example, 
In our case, the spontaneous FWM JSP (Fig~\ref{fig:SpJSP}B) is dominated by an energy-conservation trend, angle correlation with $\Delta \nu_i$-$\Delta \nu_s$ and anti-correlation %between
with $\Delta \nu_s$+$\Delta \nu_i$,
the JSP of stimulated FWM is
%\noindent However, in the Stimulated FWM case (Fig~\ref{fig:StJSP}A), the main features in the JSP appear to be 
dominated by the value of $\Delta \nu_s$, 
%value which corresponds with the seed, 
with very little noticeable change arising from $\Delta \nu_i$.
%We believe it is quite notable that the Spontaneous FWM JSP (Fig~\ref{fig:SpJSP}B) differs significantly from the Stimulated FWM JSP (Fig~\ref{fig:StJSP}A).
%Fig~\ref{fig:SpJSP} and Fig~\ref{fig:StJSP} shows that the JSP of the spontaneously emitted photon-pairs is distinctly different than the JSP of the stimulated FWM.
This follows from the
%This can be understood by noting the 
full expressions %of the
for the JSAs for the
%of a resonator for both 
spontaneous and stimulated processes in a ring resonator. % (details in the Supplementary Information). %I don't think these details are in the supplementary information actually... ref given at the end of this section but maybe it should be mentioned earlier?
The JSA of 
%a resonator for 
the SFWM process, $\Phi_{R-Sp}$, and
%an equivalent 
the JSA for the stimulated process $\Phi_{R-St}$, %can be expressed as
are given by
\begin{widetext}

\begin{align}
\Phi_{R-Sp}(\nu_s,\nu_i) =& N\int d\nu_{p1}d\nu_{p2} \alpha(\nu_{p1})\alpha(\nu_{p2})F_{p-}(\nu_{p1})F_{p-}(\nu_{p2})F^*_{s+}(\nu_s)F^*_{i+}(\nu_i) \delta(\nu_{p1} + \nu_{p2} - \nu_s - \nu_i), \label{eq:sp}\\
%%%%%
\Phi_{R-St}(\bar{\nu}_s,\nu_i) =& N^\prime\int d\nu_{p1}d\nu_{p2} \alpha(\nu_{p1})\alpha(\nu_{p2})F_{p-}(\nu_{p1})F_{p-}(\nu_{p2})F^*_{s-}(\bar{\nu}_s)F^*_{i+}(\nu_i) \delta(\nu_{p1} + \nu_{p2} - \bar{\nu}_s - \nu_i), \label{eq:st}
\end{align}

\end{widetext}
where the subscripts $p$, $s$ and $i$ represent pump, signal, and idler photons respectively, and $\alpha(\nu_p)$ is the pump field amplitude. In the first of these equations $\nu_s$ and $\nu_i$ are the frequencies of the spontaneously emitted signal and idler, while in the second $\bar{\nu}_s$ is the CW seed frequency and $\nu_i$ is the frequency of the stimulated idler; %By $\nu$, we denote the frequency of a photon, while $\bar{\nu}_s$ represents the seed frequency, corresponding to a continuous wave laser light and used for stimulating the four-wave mixing process. 
$N$ and $N^\prime$ are normalization constants, and energy conservation is expressed by the Dirac delta function, $\delta(\nu)$. The resonant field enhancement is %represented 
characterized by $F_k(\nu)$, %Assuming a Lorentzian lineshape %Energy conservation can be used to define $\nu_p^\prime = \nu_s + \nu_i - \nu_p$.
\begin{align}
    F_{k\pm}(\nu) = \frac{1}{\sqrt{L}}\frac{\gamma_k^*}{2\pi(\nu_k-\nu) \pm i\bar{\Gamma}_k},
\end{align}
where $k$ can be pump ($p$), signal ($s$) or idler ($i$), $\gamma_k$ is a coupling coefficient, $\bar{\Gamma}_k$ corresponds to the half-width half-maximum of the line-width of the resonances, and $L$ is the geometrical length of the resonator \cite{two_strategies}. 

Eqs.~\ref{eq:sp} and \ref{eq:st} are very similar; indeed, the JSIs that follow from them are identical but the JSPs are different. According to the asymptotic field model \cite{two_strategies}, the complex amplitude of the SET signal will be proportional to the JSA only when the seed field excites the system in one of its asymptotic output states. For our all-pass resonator, this would require engineering the seed in such a way that the input and the loss channel of the ring are both excited, so that the seed photons can only leave the interaction region from the through port \cite{St_JSP}. However, in the absence of direct physical access to the loss channel for our all-pass resonator we can only seed the input port; hence we are not able to excite the resonator into one of its asymptotic output state. The direct consequence is that the phase of the stimulated JSA in Eq.~\ref{eq:st} differs from the phase of the spontaneous JSA in Eq.~\ref{eq:sp} as shown in Figs.~\ref{fig:StJSP}(A) and \ref{fig:StJSP}(B). Mathematically, the resonant field enhancement factors for signal ($F^*_{s+}(\nu_s)$) in spontaneous case and seed in the stimulated case ($F^*_{s-}(\bar{\nu}_s)$) are different, leading to different JSPs. This confirms that spontaneous emission cannot be imitated completely by the corresponding stimulated emission process \cite{St_JSP, Rozema2015}. If the structure (i.e., source) generating the photon pairs is not well known and a detailed analytical model does not exist, then we require a \textcolor{red}{Q-}SpET method to extract the information from the spontaneous FWM process. 

%\textcolor{blue}{This confirmation of distinction between SET and \textcolor{red}{Q-}SpET has not been predicted by the recent experiments \cite{Thekkadath2022, Phase_tomography} on acquiring the JSP\textcolor{red}{, ruling out its use as a general solution}}.Recent results \cite{St_JSP} predicted that SET alone is not sufficient to extract JSP of a resonant photon-pair source. The correct JSP can only be extracted from SET measurements by using a perfect analytical model of the photon-pair generation process for that source\textcolor{red}{, anything less would constitute an approximation.}
%If an analytical model of the photon-pair generation insidea resonant source exists, it may be used to transform the SET measurements to the correct JSP..0

Among all the demonstrations of JSP, the most direct experimental method is by Jizan et al. \cite{Phase_tomography} which uses the phase-sensitive nature of the SET method, alongside fast and high-resolution classical interferometers that reduce the complexity posed by the other methods \cite{Thekkadath2022, MacLean2019}. 
%The beauty of our method is that it gives similar expression (Eqs.~\ref{eq:probability} and \ref{eq:JSP}) as for SET and thus the measurement is as direct as \cite{Phase_tomography}, except, we use quantum interference.
Q-SpET's approach results in surprisingly similar expressions for the stimulated and spontaneous processes (Eqs.~\ref{eq:probability} and \ref{eq:JSP}), exploiting quantum interference to directly measure the JSP as in phase-sensitive SET \cite{Phase_tomography}. This allows us to extract the correct JSP at the expense of the lower detection rate imposed by the use of single-photon detectors. Our method also allows us to eliminate any complex phase retrieval algorithms, Fourier filtering, or time-resolved single-photon counting spectrometers (TRSPSs) \cite{Thekkadath2022, MacLean2018, MacLean2019, Tischler2015}. Our interferometry also benefits from the inherent phase stability of the integrated photonic platform, and due to our implementation, does not require any fast modulators \cite{Davis2020}.

%Our method also does not require complex phase retrieval algorithms, Fourier filtering, and time-resolved single-photon counting spectrometer (TRSPS) \cite{Thekkadath2022, MacLean2018, MacLean2019, Tischler2015}.}

%\textcolor{red}{
%It is unclear if alternative methods \cite{Thekkadath2022, MacLean2019} can extract the correct JSP of a resonant photon-pair source, but at a minimum they cannot alleviate experimental constraints by moving to the stimulated regime. 
%Nevertheless, it is possible to extract the correct JSP using a self-referenced method \cite{Davis2020} by implementing Electro-Optic Shearing Interferometry as well as additional post-processing using Fourier filtering, a rather more complex approach than that presented here.}
%We have successfully shown that our SpET method can extract JSP of a conventional micro-resonator photon-pair source, and our method can potentially be used for more complex resonant sources.
%Although \cite{Davis2020} is a self-referenced method and possibly extract the correct JSP of a resonant photon-pair source,   }

%%%%%%%%%%%%%%%%%%%%%%%%%%%%%%%%%%%%%%%%%%%%%%%%
%Also, our method is intrinsically phase stable, and does not require detailed prior knowledge the pump, nor any complex computational algorithms or modelling. -- point to re-address.
%Conclusion\\
\section{Conclusion and Outlook}
%In summary, we have performed the first direct measurement of the joint spectral phase of a biphoton wave function without the need of a prior model of the photon source,
In summary, Q-SpET is a precise and resource-efficient technique in both detection and processing overhead, that recovers the joint spectral phase of a biphoton wave function without relying on an analytical model of the photon source, by using quantum interference between biphoton wave functions from two different sources. We have demonstrated this method using a photonic chip consisting of an optical resonator structure and a reference waveguide structure --- the source of our flat JSP. The interference we observe is intrinsically phase stable and, beyond typical spectral characteristics, does not require any other knowledge of the pump. Our method shows for the first time a clear distinction between the spontaneous and stimulated four-wave mixing processes within the same photonic device, and over the same wavelength range, by comparing the measured JSPs.

This concept can be further applied to any unknown source, as long as its biphoton wavefunction can be interfered with that of a well-characterized reference. Furthermore, this concept is not restricted to photon-pair sources and may be used for any quantum state generated in superposition using, for instance, the weak pump regime. % Since SpET could be used to certify a quantum process, a potential application could be to secure a communication channel by sharing the resultant state ($|\psi_{3}\rangle$, Fig.~\ref{fig:chip}) between two parties. Recipients of state $|\psi_{3}\rangle$ could use SpET to recover the JSP, and verify that the state they are receiving has not been tampered with by a third-party. Here, the channel is effectively encoded using the nonlinear process that created $|\psi_{3}\rangle$.
%We also envisage this method to be adaptable for the characterization of other light-emission processes such as the emission from quantum dots or molecules. 
Additionally, JSPs of the spontaneous emission of a source could be used as signatures for different sources, effectively enabling certification protocols. Overall, the method presented here opens up possibilities in future metrology and tomographic methods, applications in calibration protocols and the definition of better standards for quantum information applications.
\newpage

%%%%%%%%%%%%%%%%%%%%%%%%%%%%%%%%%%%%%%%%%%%%%%%%
\bibliographystyle{unsrt}
\bibliography{references}

\begin{thebibliography}{10}

\bibitem{Ekert91}
Artur~K. Ekert.
\newblock Quantum cryptography based on bell's theorem.
\newblock {\em Physical Review Letter}, 67:661--663, Aug 1991.

\bibitem{Micius_entanglement}
Juan Yin, Yuan Cao, Yu-Huai Li, Sheng-Kai Liao, Liang Zhang, Ji-Gang Ren, Wen-Qi Cai, Wei-Yue Liu, Bo~Li, Hui Dai, et~al.
\newblock Satellite-based entanglement distribution over 1200 kilometers.
\newblock {\em Science}, 356(6343):1140--1144, 2017.

\bibitem{SKJoshi2018}
Siddarth~Koduru Joshi, Djeylan Aktas, Sören Wengerowsky, Martin Lončarić, Sebastian~Philipp Neumann, Bo~Liu, Thomas Scheidl, Guillermo~Currás Lorenzo, Željko Samec, Laurent Kling, Alex Qiu, Mohsen Razavi, Mario Stipčević, John~G. Rarity, and Rupert Ursin.
\newblock A trusted node-free eight-user metropolitan quantum communication network.
\newblock {\em Science Advances}, 6(36):eaba0959, 2020.

\bibitem{Multi_dimension}
Jianwei Wang, Stefano Paesani, Raffaele Ding, Yunhong~Santagati, Paul Skrzypczyk, Alexia Salavrakos, Jordi Tura, Remigiusz Augusiak, Laura Mančinska, Davide Bacco, Damien Bonneau, Joshua~W. Silverstone, Qihuang Gong, Antonio Acín, Karsten Rottwitt, Jeremy~L. Oxenlowe, Leif K.and~O’Brien, Anthony Laing, and Mark~G. Thompson.
\newblock Multidimensional quantum entanglement with large-scale integrated optics.
\newblock {\em Science}, 360:285--291, 2018.

\bibitem{RevHom2}
J.~W. Silverstone, D.~Bonneau, K.~Ohira, N.~Suzuki, H.~Yoshida, N.~Iizuka, M.~Ezaki, C.~M. Natarajan, M.~G. Tanner, R.~H. Hadfield, V.~Zwiller, G.~D. Marshall, J.~G. Rarity, J.~L. O'Brien, and M.~G. Thompson.
\newblock Silicon-on-insulator integrated source of polarization-entangled photons.
\newblock {\em Nature Photonics}, 8:104--108, 2014.

\bibitem{Metrology_review}
Paul-Antoine Moreau, Ermes Toninelli, Thomas Gregory, and Miles~J Padgett.
\newblock Imaging with quantum states of light.
\newblock {\em Nature Reviews Physics}, 1:367--380, 2019.

\bibitem{Lemos2014}
Gabriela~Barreto Lemos, Victoria Borish, Garrett~D. Cole, Sven Ramelow, Radek Lapkiewicz, and Anton Zeilinger.
\newblock Quantum imaging with undetected photons.
\newblock {\em Nature}, 512:409--412, 2014.

\bibitem{LOQC}
E.~Knill, R.~Laflamme, and G.~J. Milburn.
\newblock A scheme for efficient quantum computation with linear optics.
\newblock {\em Nature}, 409:46--52, 2001.

\bibitem{Rudolph2017}
Terry Rudolph.
\newblock {Why I am optimistic about the photonic route to quantum computing}.
\newblock {\em APL Photonics}, 2:030901, 2017.

\bibitem{Gentile2021}
Antonio~A. Gentile, Brian Flynn, Sebastian Knauer, Nathan Wiebe, Stefano Paesani, John~G. Granade, Christopher E.and~Rarity, Raffaele Santagati, and Anthony Laing.
\newblock A scheme for efficient quantum computation with linear optics.
\newblock {\em Nature Physics}, 17:837--843, 2021.

\bibitem{pittman2005heralding}
TB~Pittman, BC~Jacobs, and JD~Franson.
\newblock Heralding single photons from pulsed parametric down-conversion.
\newblock {\em Optics communications}, 246(4-6):545--550, 2005.

\bibitem{Kim2005}
Yoon-Ho Kim and Warren~P. Grice.
\newblock Measurement of the spectral properties of the two-photon state generated via type ii spontaneous parametric downconversion.
\newblock {\em Opt. Lett.}, 30(8):908--910, Apr 2005.

\bibitem{Liscidini2013}
M.~Liscidini and J.~E. Sipe.
\newblock Stimulated emission tomography.
\newblock {\em Physical Review Letters}, 111:193602, 2013.

\bibitem{Grassani2016}
Davide Grassani, Angelica Simbula, Stefano Pirotta, Matteo Galli, Matteo Menotti, Nicholas~C Harris, Tom Baehr-Jones, Michael Hochberg, Christophe Galland, Marco Liscidini, and Daniele Bajoni.
\newblock {Energy correlations of photon pairs generated by a silicon microring resonator probed by Stimulated Four Wave Mixing}.
\newblock {\em Scientific reports}, 6(March):23564, 2016.

\bibitem{St_JSP}
Massimo Borghi.
\newblock Phase-resolved joint spectra tomography of a ring resonator photon pair source using a silicon photonic chip.
\newblock {\em Optics Express}, 28(5):7442--7462, Mar 2020.

\bibitem{Mittal2021}
Sunil Mittal, Venkata~Vikram Orre, Elizabeth~A. Goldschmidt, and Mohammad Hafezi.
\newblock {Tunable quantum interference using a topological source of indistinguishable photon pairs}.
\newblock {\em Nature Photonics}, 15(7):542--548, 2021.

\bibitem{davis2018measuring}
Alex~OC Davis, Val{\'e}rian Thiel, Micha{\l} Karpi{\'n}ski, and Brian~J Smith.
\newblock Measuring the single-photon temporal-spectral wave function.
\newblock {\em Physical review letters}, 121(8):083602, 2018.

\bibitem{Davis2020}
Alex O.~C. Davis, Val\'{e}rian Thiel, and Brian~J. Smith.
\newblock Measuring the quantum state of a photon pair entangled in frequency and time.
\newblock {\em Optica}, 7(10):1317--1322, Oct 2020.

\bibitem{beduini2014interferometric}
Federica~A Beduini, Joanna~A Zieli{\'n}ska, Vito~G Lucivero, Yannick~A de~Icaza~Astiz, and Morgan~W Mitchell.
\newblock Interferometric measurement of the biphoton wave function.
\newblock {\em Physical review letters}, 113(18):183602, 2014.

\bibitem{MacLean2019}
Jean-Philippe~W. MacLean, Sacha Schwarz, and Kevin~J. Resch.
\newblock Reconstructing ultrafast energy-time-entangled two-photon pulses.
\newblock {\em Phys. Rev. A}, 100:033834, Sep 2019.

\bibitem{Tischler2015}
N.~Tischler, A.~B\"use, L.~G. Helt, M.~L. Juan, N.~Piro, J.~Ghosh, M.~J. Steel, and G.~Molina-Terriza.
\newblock Measurement and shaping of biphoton spectral wave functions.
\newblock {\em Phys. Rev. Lett.}, 115:193602, Nov 2015.

\bibitem{Thekkadath2022}
G.~S. Thekkadath, B.~A. Bell, R.~B. Patel, M.~S. Kim, and I.~A. Walmsley.
\newblock Measuring the joint spectral mode of photon pairs using intensity interferometry.
\newblock {\em Phys. Rev. Lett.}, 128:023601, Jan 2022.

\bibitem{HOM}
C.~K. Hong, Z.~Y. Ou, and L.~Mandel.
\newblock Measurement of subpicosecond time intervals between two photons by interference.
\newblock {\em Physical Review Letter}, 59:2044--2046, Nov 1987.

\bibitem{Wang2018}
Xi-lin Wang, Yi-han Luo, He-liang Huang, Ming-Cheng Chen, Zu-en Su, Chang Liu, Chao Chen, Wei Li, Yu-Qiang Fang, Xiao Jiang, Jun Zhang, Li~Li, Nai-Le Liu, Chao-Yang Lu, and Jian-Wei Pan.
\newblock {18-Qubit Entanglement with Six Photons’ Three Degrees of Freedom}.
\newblock {\em Physical Review Letters}, 120:260502, 2018.

\bibitem{Ekert1992}
Artur~K. Ekert, John~G. Rarity, Paul~R. Tapster, and G.~Massimo Palma.
\newblock {Practical quantum cryptography based on two-photon interferometry}.
\newblock {\em Physical Review Letters}, 69:1293, 1992.

\bibitem{NOON_theory}
Pieter Kok, Hwang Lee, and Jonathan~P. Dowling.
\newblock Creation of large-photon-number path entanglement conditioned on photodetection.
\newblock {\em Physical Review A}, 65:052104, Apr 2002.

\bibitem{Fukuda:05}
Hiroshi Fukuda, Koji Yamada, Tetsufumi Shoji, Mitsutoshi Takahashi, Tai Tsuchizawa, Toshifumi Watanabe, Jun ichi Takahashi, and Sei ichi Itabashi.
\newblock Four-wave mixing in silicon wire waveguides.
\newblock {\em Optics Express}, 13(12):4629--4637, Jun 2005.

\bibitem{Jeffrey2004}
Evan Jeffrey, Nicholas~A Peters, and Paul~G Kwiat.
\newblock Towards a periodic deterministic source of arbitrary single-photon states.
\newblock {\em New Journal of Physics}, 6:100--100, jul 2004.

\bibitem{Migdall2002}
A.~L. Migdall, D.~Branning, and S.~Castelletto.
\newblock {Tailoring single-photon and multiphoton probabilities of a single-photon on-demand source}.
\newblock {\em Physical Review A - Atomic, Molecular, and Optical Physics}, 66(5):4, 2002.

\bibitem{Faruque2018}
Imad~I. Faruque, Gary~F. Sinclair, Damien Bonneau, John~G. Rarity, and Mark~G. Thompson.
\newblock {On-chip quantum interference with heralded photons from two independent micro-ring resonator sources in silicon photonics}.
\newblock {\em Optics Express}, 26(16):20379--20395, 2018.

\bibitem{LLewellyn2020}
Daniel Llewellyn, Yunhong Ding, Imad~I. Faruque, Stefano Paesani, Davide Bacco, Raffaele Santagati, Yan-Jun Qian, Yan Li, Yun-Feng Xiao, Marcus Huber, Mehul Malik, Gary~F. Sinclair, Xiaoqi Zhou, Karsten Rottwitt, Jeremy~L. O’Brien, John~G. Rarity, Qihuang Gong, Leif~K. Oxenlowe, Jianwei Wang, and Mark~G. Thompson.
\newblock Chip-to-chip quantum teleportation and multi-photon entanglement in silicon.
\newblock {\em Nature Physics}, 16:148–153, 2020.

\bibitem{zhang2021squeezed}
Y~Zhang, M~Menotti, K~Tan, VD~Vaidya, DH~Mahler, LG~Helt, L~Zatti, M~Liscidini, B~Morrison, and Z~Vernon.
\newblock Squeezed light from a nanophotonic molecule.
\newblock {\em Nature communications}, 12(1):1--6, 2021.

\bibitem{Phase_tomography}
Iman Jizan, Bryn Bell, L.~G. Helt, Alvaro~Casas Bedoya, Chunle Xiong, and Benjamin~J. Eggleton.
\newblock Phase-sensitive tomography of the joint spectral amplitude of photon pair sources.
\newblock {\em Optics Letter}, 41(20):4803--4806, Oct 2016.

\bibitem{Bi_photon}
Zhenshan Yang, Marco Liscidini, and J.~E. Sipe.
\newblock Spontaneous parametric down-conversion in waveguides: A backward heisenberg picture approach.
\newblock {\em Physical Review A}, 77:033808, Mar 2008.

\bibitem{RevHom1}
Laurent Olislager, Jassem Safioui, St\'{e}phane Clemmen, Kien~Phan Huy, Wim Bogaerts, Roel Baets, Philippe Emplit, and Serge Massar.
\newblock Silicon-on-insulator integrated source of polarization-entangled photons.
\newblock {\em Optics Letter}, 38(11):1960--1962, Jun 2013.

\bibitem{two_strategies}
Milica Banic, Luca Zatti, Marco Liscidini, and J.~E. Sipe.
\newblock Modeling nonlinear optics in lossy microring systems: two strategies.
\newblock {\em arXiv:2111.14711}, 2021.

\bibitem{Faruque2019}
Imad~I. Faruque, Gary~F. Sinclair, Damien Bonneau, Takafumi Ono, Christine Silberhorn, Mark~G. Thompson, and John~G. Rarity.
\newblock {Estimating the Indistinguishability of Heralded Single Photons Using Second-Order Correlation}.
\newblock {\em Physical Review Applied}, 12:054029, 2019.

\bibitem{Rozema2015}
Lee~A. Rozema, Chao Wang, Dylan~H. Mahler, Alex Hayat, Aephraim~M. Steinberg, John~E. Sipe, and Marco Liscidini.
\newblock Characterizing an entangled-photon source with classical detectors and measurements.
\newblock {\em Optica}, 2(5):430--433, May 2015.

\bibitem{MacLean2018}
Jean-Philippe~W. MacLean, John~M. Donohue, and Kevin~J. Resch.
\newblock Direct characterization of ultrafast energy-time entangled photon pairs.
\newblock {\em Phys. Rev. Lett.}, 120:053601, Jan 2018.

\bibitem{Wilkes:16}
C.~M. Wilkes, X.~Qiang, J.~Wang, R.~Santagati, S.~Paesani, X.~Zhou, D.~A.~B. Miller, G.~D. Marshall, M.~G. Thompson, and J.~L. O'Brien.
\newblock 60 db high-extinction auto-configured mach--zehnder interferometer.
\newblock {\em Optics Letter}, 41(22):5318--5321, Nov 2016.

\bibitem{Walls_Milburn}
D.F. Walls and Gerard~J. Milburn.
\newblock {\em Quantum Optics}.
\newblock Springer Link, 2008.

\bibitem{Vernon_lossy_resonators}
Z.~Vernon and J.~E. Sipe.
\newblock Spontaneous four-wave mixing in lossy microring resonators.
\newblock {\em Phys. Rev. A}, 91:053802, May 2015.

\end{thebibliography}
%%%%%%%%%%%%%%%%%%%%%%%%%%%%%%%%%%%%%%%%%%%%%%%%
\clearpage
\newpage
\section*{Methods}
\textbf{Experimental setup}: Figure~\ref{fig:chip} shows the photonic chip and Fig~\ref{fig:setup} shows the optical setup. Our chip is designed and fabricated on a conventional silicon-on-insulator (SOI) platform with waveguide cross-section 220~nm $\times$ 500~nm for single-mode operation. Photon-pairs are generated through SFWM, where two pump photons are converted into a pair of photons with different frequencies, commonly denoted as signal and idler photons. We use a pulsed tuneable pump laser centered at a wavelength of 1550~nm (PriTel FFL 50~MHz, 15~ps, 6~mW) and pre-filtered with two broadband filters (BFs). This is a crucial step to reduce the pump laser's background contribution at SFWM signal-idler wavelengths of interest with nearly 120~dB attenuation. These broadband filters are commercially available Dense Wavelength Division Multiplexers from Opneti. The intensity and polarization of the pump are then adjusted using a variable optical attenuator (VOA) and the polarization controller. The polarization controller orients the light to a TE polarization for optimal coupling to the onc-hip vertical grating coupler (VGC). The measured coupling loss of these VGC is -4.5~dB. The pump is then distributed across two separate on-chip photonic SFWM sources (spiral length of 2.8~mm, resonator geometrical length of 355~$\mu$m), and the residual pump filtered out by asymmetric MZIs \cite{Wilkes:16}. The generated photon-pairs propagate through the circuit and are interfered using an on-chip Mach-Zehnder Interferometer (MZI) structure. The output is then coupled back into a single-mode fiber (SMF28) using VGCs, and filtered using BFs, before being collected by single-photon detectors. The detectors are connected to a time-interval analyzer to record the coincidence counts used to produce the interferograms.

Our resonator (Fig.~\ref{fig:chip}) resembles a racetrack resonator except the coupling region, which usually is implemented by a directional coupler. Instead, this resonator has a Mach-Zehnder interferometer which acts as a variable coupling region. By applying a voltage to the phase element in this interferometer, we can vary the resonance linewidth. This is shown in Fig.~\ref{fig:res}, where the resonance can be tuned from over-coupled to critically-coupled to under-coupled regions. We have chosen an over-coupled resonance with a reasonable quality factor. This allows relatively high resonance enhancement while allowing broad enough linewidth (FWHM 60~pm) to collect enough data points within the resonance using our on-chip filters.

For SET, we use a narrow linewidth (125~MHz or 1~pm) tuneable continuous wave laser (Yenista) to seed to the FWM process at the signal wavelength. The idler photons generated by StFWM are interfered, coupled out of the chip and filtered (BF) to remove any background pump. We use a Finisar Optica Spectrum Analyzer (OSA) with 20~MHz or 0.16~pm resolution to record the spectra of the idler.

\textbf{Data processing}: The experimentally recorded interferograms for spontaneous emission tomography have been passed through Gaussian and moving average filters before using them for estimating the JSP. This filtering essentially removes any high frequency noise present in the interferograms.

\section*{Data availability}
The data used to produce the plots within this article are available at ...

\section*{Code availability}
The MATLAB program used for processing the data is nevertheless available from the corresponding author on reasonable request.

\section*{Acknowledgements}

\section*{Author contributions}
Authors I.I.F and B.M.B contributed equally to this work. I.I.F. conceived the idea and method of the spontaneous emission tomography. I.I.F. and M.B. designed the photonic circuit. B.M.B. and I.I.F. performed the experiment. B.M.B., I.I.F. and J.B. analyzed the data and performed the simulation. J.B. and J.G.R supervised the project. All the authors contributed in writing the manuscript. 
% based on the interferometric scheme and the idea reported in \cite{St_JSP, Phase_tomography}

\section*{Funding}
This work was supported by the Engineering and Physical Sciences Research Council (EPSRC) under the grant EP/L0240201. B.M.B. acknowledges the support of the EPSRC training grant EP/LO15730/1. The authors also acknowledge EPSRC's additional support of the work via grants EP/M024458/1, EP/K033085/1 and EP/N015126/1.

\section*{Competing interests}
The authors declare no competing interests.

% \section*{Extended data}
% \renewcommand\thefigure{Extended Data \arabic{figure}}    
% \setcounter{figure}{0}

% \begin{figure*}[h!]
% \centering
% \includegraphics[width=\textwidth]{FigS2_fringe.pdf}
% \caption{Using the experimental setup of the SpET we can observe the following interaction among the photon-pairs from the waveguide and the photon-pairs. \textbf{A}. The top left graph (i) shows a reference classical interference fringe, the remaining graphs (ii, iii, and iv) show quantum interference fringes measured for different signal-idler frequency points - as mapped in {C}%, which has a distinctive doubling of the period - despite the remarkable difference between the waveguide and the resonator
% . \textbf{B}. Simulated JSAs of the waveguide (left) and the resonator (right) on-chip sources based on the measured parameters.
% }
% \label{fig:qi}
% \end{figure*}

%\end{multicols}

%%%%%%%%%%%%%%%%%%%%%%%%%
\clearpage
\newpage

\onecolumngrid

\section*{Supplemental Material: Spontaneous emission tomography}

\renewcommand\thefigure{S\arabic{figure}}    
\setcounter{figure}{0}    
\renewcommand{\thetable}{S\arabic{table}} 
\setcounter{table}{0}
\renewcommand{\theequation}{S\arabic{equation}} 
\setcounter{equation}{0}

\section{Experimental setup}
\label{Supp:setup}
\begin{figure*}[h!]
\centering
\includegraphics[width=\textwidth]{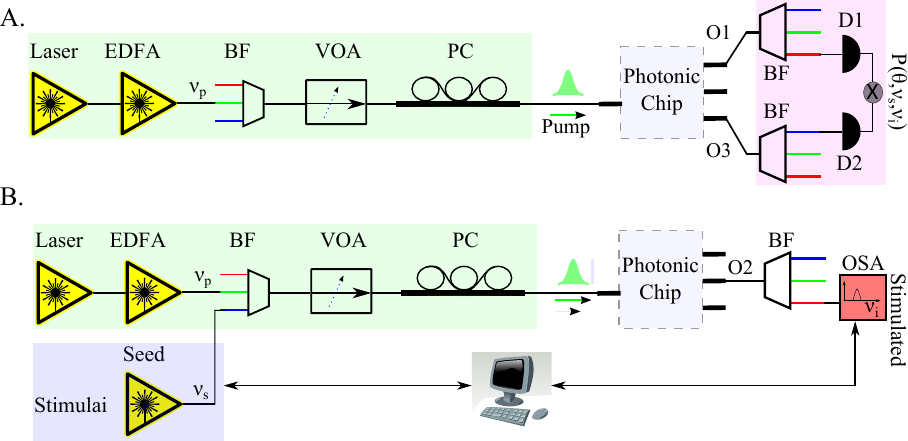}
\caption{Experimental setup. \textbf{A} Spontaneous emission tomography. Schematic of the experimental setup for spontaneous emission tomography. Signal and idler photons are collected from the on-chip filters at O1 and O3 and filtered again using off chip broadband filters (BFs) to remove background pump photons and detected using single-photon detectors (D1 and D2) for coincidence counts. \textbf{B} Stimulated emission tomography. Schematic of the experimental setup used for stimulated emission tomography. Similarly to the set-up presented in Fig~\ref{fig:SpJSP} a pulsed laser is used as a pump and a second (seed) laser is also added. In this case an Optical Spectrum Analyzer (OSA) is used to measure the stimulated idler spectra.
}
\label{fig:setup}
\end{figure*}

\begin{figure}[h!]
\centering
\includegraphics[scale = 0.93]{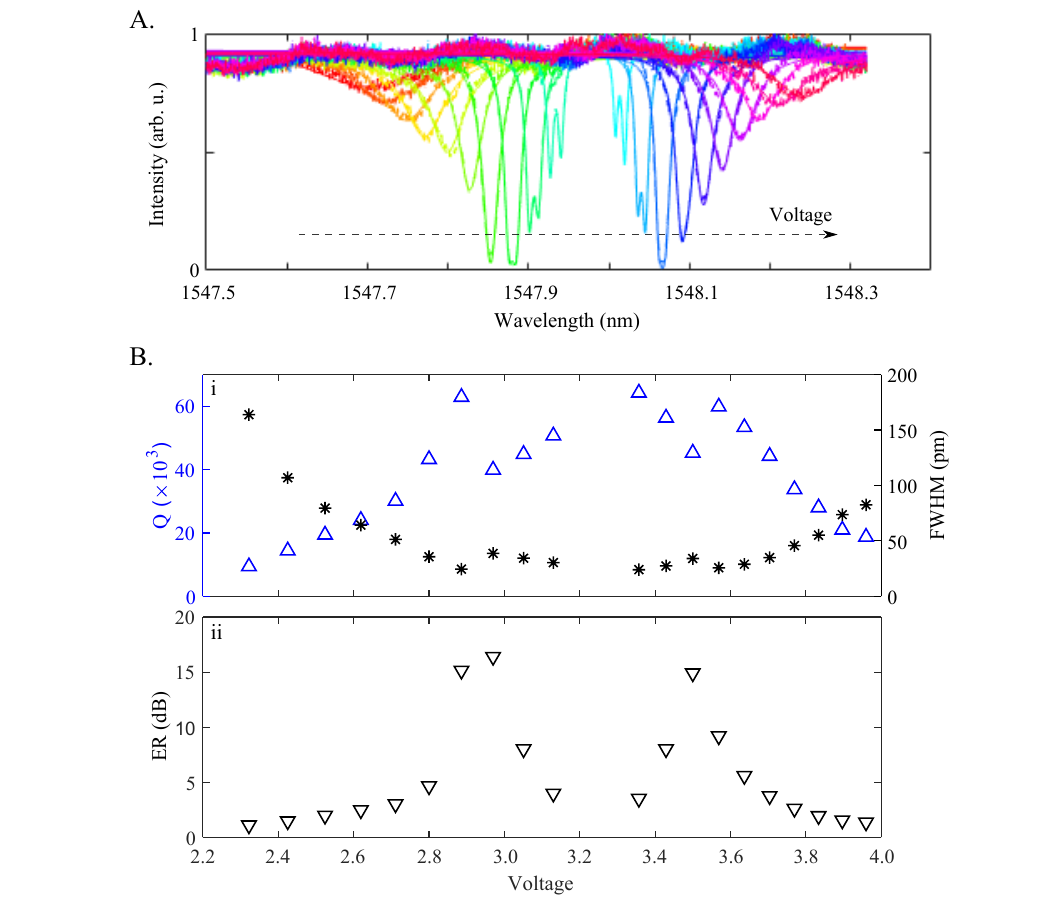}
\caption{Tuneable coupling of our resonator. \textbf{A}. Resonant lineshapes as a function of applied voltage in the coupling region. The applied voltage tunes the fraction of light that is being coupled to the resonator. Thus, the coupling can be tuned from over-coupling to critical-coupling to under-coupling. \textbf{B}. Extracted parameters of the resonances as a function of applied voltage. i) Quality factor (Q) and full-width at half-maximum (FWHM) of the linewidth vs voltage. ii) extinction ratio (ER) vs voltage. The highest extinction configurations are the critically-coupled configurations. We chose 60 pm linewidth---an over-coupled configuration.}
\label{fig:res}
\end{figure}

%\vspace{-15mm}
As shown in Fig.~\ref{fig:setup}, we use a photonic chip for our experiment with an off-chip pump laser and two off-chip single-photon detectors. The details of our design parameters, simulation parameters and experimental values are listed in Tab.~\ref{tab:parameters}. The chosen linewidth, Full-Width at Half-Maximum (FWHM), of the resonator is 7.5 GHz (or 60 pm), which is an over-coupled configuration as shown in Fig.~\ref{fig:res}. Subsequently its four-wave mixing bandwidth is considerably narrower than that of the waveguide source ($>$10 nm) \cite{Faruque2019}. This can be seen from the JSA comparisons in Fig.~\ref{fig:qi} B.
%%%%%%%%
% \enlargethispage{-5.0cm}
% \noindent\begin{picture}(0,0)
% \put(0,0){
% \begin{minipage}{\textwidth}
% \centering
% \vspace{0mm}
% \includegraphics[width=0.99\textwidth]{FigS1_Setup.pdf}
% \captionof{figure}{Experimental setup. \textbf{A} Spontaneous emission tomography. Schematic of the experimental setup for spontaneous emission tomography. Signal and idler photons are collected from the on-chip filters at O1 and O3 and filtered again using off chip broadband filters (BFs) to remove background pump photons and detected using single-photon detectors (D1 and D2) for coincidence counts. \textbf{B} Stimulated emission tomography. Schematic of the experimental setup used for stimulated emission tomography. Similarly to the set-up presented in Fig~\ref{fig:SpJSP} a pulsed laser is used as a pump and a second (seed) laser is also added. In this case an Optical Spectrum Analyser (OSA) is used to measure the stimulated idler spectra.
% }
% \label{fig:setup}
% \end{minipage}
% }
% \end{picture}%
%%%%%% 
%%%%%% table
\begin{table}[h!]
\centering
\begin{tabular}{ p{5.5cm}|p{4.5cm}} 
 \hline
 Parameters & Values \\ 
 \hline
Waveguide dimension & 220 nm $\times$ 500 nm\\ 
Waveguide length & 2.8 mm \\
Waveguide loss & 2.5 dB/cm, 0.01 dB/bend \\
Compact model, [$n_1$, $n_2$, $n_3$] &
[2.4473, -1.1327, -0.0440] \\
Group velocity, ng & 4.181 at 1.55$\mu$m \\
Resonator linewidth & 7.44 GHz (59.6 pm) \\
Resonator geometrical length & 363 $\mu$m \\
On-chip filter linewidth & 1.25 GHz (10 pm) \\
Pump wavelength & 1546.23 nm \\
Pump repetition rate & 50 MHz \\
Pump linewidth & 15 ps \\
Signal wavelength & 1538.55 nm \\
Idler wavelength & 1553.98 nm \\
 \hline
\end{tabular}
\caption{Compact model: $n_{eff}=n_1+n_2(\lambda-\lambda_0)+n_3(\lambda-\lambda_0)^2$, where $\lambda$ is wavelength in $\mu$m.}
\label{tab:parameters}
\end{table}
%%%%%%%%%%%%%%%%%%%
%%%%%%%%%%%%%%%%%%%%%%%
%%%%%%%%%%%%%%%%%%%%%%%%
%\vspace{-50mm}
%%%%%%%%%%%%%%%%%%%%%%%%%%%%%%%%%%%%%%
\section{Quantum interference between two different sources}
Prior experiments show that near-unity visibility (fidelity) can be achieved when photons from the same source or identical sources are interfered \cite{RevHom1, RevHom2}, as signal and idler photons have the same JSA.
However, our two sources have different JSAs, leading to measurable visibility differences following the interference.
%%%%%%%%%%%%%%%%%%%%%%%%%%%%%%%%%%%%%%
\begin{figure*}[h!]
\centering
\includegraphics[width=\textwidth]{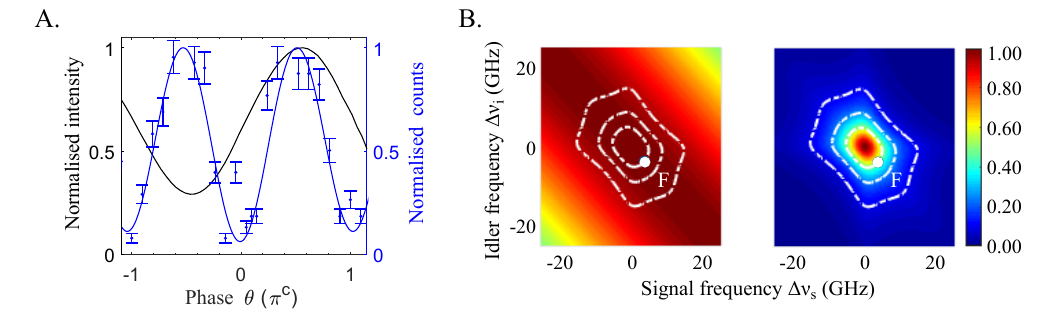}
\caption{Quantum interference between two dissimilar sources.
Using the experimental setup of the \textcolor{red}{Q-}SpET we can observe the following interaction among the photon-pairs from the waveguide and the photon-pairs. \textbf{A}. The plot shows reference classical interference fringe (black), and quantum interference data (blue) and a fitted fringe (blue). The position of the on-chip spectral filter is marked by a white dot in {B}. The quantum interference fringe has a distinctive doubling of the period and a visibility of $~88\%$. \textbf{B}. Simulated JSAs of the waveguide (left) and the resonator (right) on-chip sources based on the measured parameters. The white contours lines show constant amplitude of JSA. From innermost to outer contours, the amplitudes are 50\%, 75\% and 90\%. This plot shows that withing the range of the JSA of a resonator, the waveguide JSA varies very little.
}
\label{fig:qi}
\end{figure*}
%%%%%%%%%%%%%%%%%%%%%%%%%%%%%%%%%%%%%%
%%%%%%%%%%%%%%%%%%%%%%%%%%%%%%%%%%%%%%
\begin{figure*}[h!]
\centering
\includegraphics[width=0.8\textwidth]{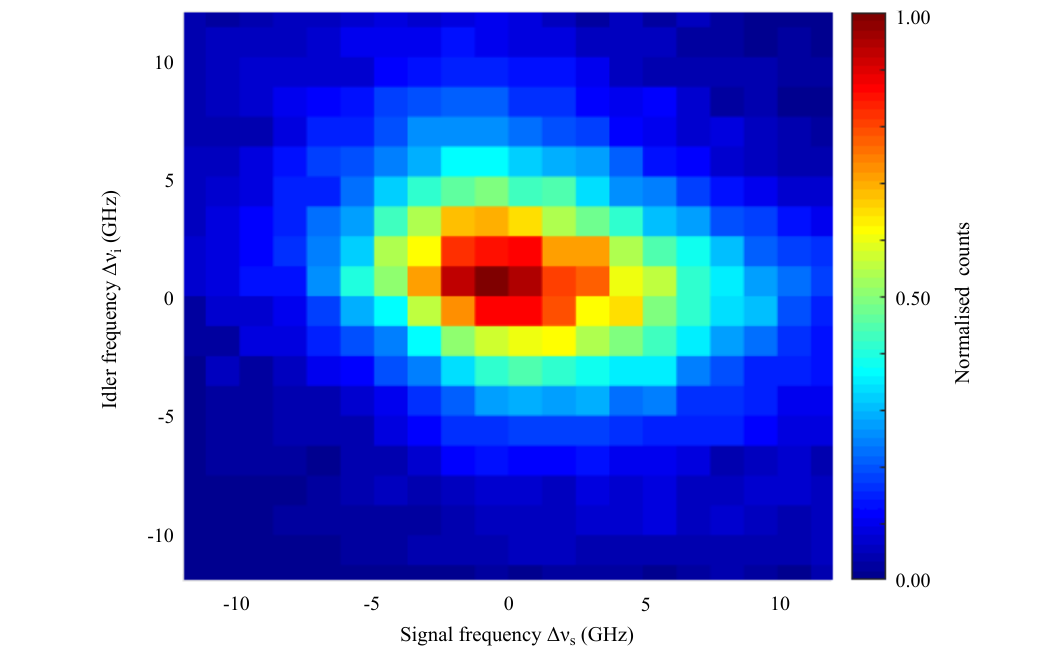}
\caption{Joint spectral intensity from spontaneous photon-pair emission. We have only pumped the micro-resonator in our circuit and used the setup of Fig.~\ref{fig:setup}A to collect the signal and idler photons to construct the JSI.
}
\label{fig:SpJSI}
\end{figure*}
%%%%%%%%%%%%%%%%%%%%%%%%%%%%%%%%%%%%%%
%For example, in Fig~\ref{fig:qi}B-i. the top left graph shows the measured interference fringe of a classical state. The remaining graphs (ii, iii, and iv) show the quantum interference fringes obtained by scanning $\theta$. The exact values used in each case are mapped within the JSAs of each separate source as presented in Fig~\ref{fig:qi}C. Specifically, Fig~\ref{fig:qi}B-ii, corresponds to the case where on-chip filters were used to collect photons only from centre of the JSA of two sources while Fig~\ref{fig:qi}B-iii and Fig~\ref{fig:qi}B-iv correspond to measurements with an offset on the anti correlated energy conservation line (iii) or or correlated momentum conservation line (iv).
Fig.~\ref{fig:qi} A, shows a quantum interference fringe between our sources. A classical interference fringes is also plotted here for comparison. The white dot, labeled as ``F", in Fig.~\ref{fig:qi} B, represents the position of the on-chip tuneable filter (10 pm bandwidth) while collecting the data of this quantum interference fringe.
As expected, the quantum interference has the distinctive period-doubling compared to the classical case, with a measurable visibility of about 88\%, which is remarkable considering these are two different sources.
%%%%%
%%%%%%%%%%%%%%%%%%%%%%%%%%%%%%%%%%%%%%%%
\section{Tomography model and effect of loss}
\label{Supp:maths}
This section elaborates Eqs.~\ref{eq:pump_split}-\ref{eq:JSP} and the analytical model used for estimating the JSP. %These phenomena are present to some degree in practice; here we consider their effects on our tomography.

%\subsection{Tomography model}

\begin{figure}[h!]
\centering
\includegraphics[width=\textwidth]{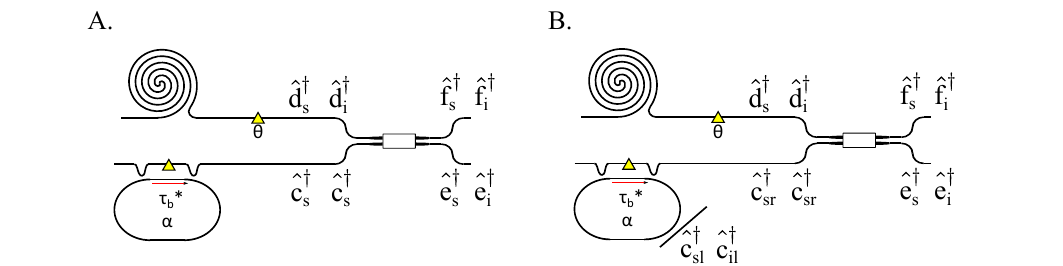}
\caption{Effect of loss represented by a phantom channel. \textbf{A}. Beam splitter transformation operators at the final 50:50 beam splitter in the circuit. \textbf{B}. Phantom channels representing loss during the photon-pair generation process in the resonator. The relevant operators for different parts of the circuit are indicated at the relevant parts of the circuit.
}
\label{fig:phantom}
\end{figure}

Neglecting scattering loss for the moment, we can write the quantum state of the system after the pump filters as
\begin{align}
|\Psi_2\rangle = \int d\nu_s d\nu_i \left(N_R\Phi_R \left(\nu_s, \nu_i \right) c^{\dagger}_s(\nu_s) c^{\dagger}_i(\nu_i) e^{j2\theta} + N_W \Phi_W\left(\nu_s, \nu_i \right) d^{\dagger}_s(\nu_s) d^{\dagger}_i(\nu_i) \right) |\mathrm{vac}\rangle. \label{eq:no_phantom}
\end{align}
The operators $c_k(\nu_k)$ and $d_k(\nu_k)$ refer to the parts of the circuit associated with the resonant and reference sources, respectively (see Fig.~\ref{fig:phantom}). During the interference at the final 50:50 beam splitter (Fig.~\ref{fig:phantom}), these operators transform according to 
\begin{align}
c_{k}^{\dagger}(\nu_k) = \frac{1}{\sqrt{2}}\left( e^{\dagger}_k(\nu_k) + i f^{\dagger}_k(\nu_k) \right), \label{eq:BSc}\\
d_{k}^{\dagger}(\nu_k) = \frac{1}{\sqrt{2}}\left( i e^{\dagger}_k(\nu_k) + f^{\dagger}_k(\nu_k) \right), \label{eq:BSd}
\end{align}
where ($k=\{s,i\}$) and the operators $e_k(\nu_k)$ and $f_k(\nu_k)$ refer to the output arms of the interferometer \cite{Walls_Milburn}.

%%%%%%%%%%%%%%%%
Using $\hat{n}_s$ and $\hat{n}_i$ to denote the number operators for signal and idler photons in the ``$f$" arm of the interferometer, the coincidence counts in this arm are expressed as
\begin{align}
P\left( \theta, \eta, \nu_s, \nu_i \right) = \langle \hat{n}_s \hat{n}_i \rangle = \frac{1}{4} \left| N_R \Phi_R\left(\nu_s, \nu_i \right) e^{i2\theta} +  N_W \Phi_W\left(\nu_s, \nu_i \right) \right|^2. \label{eq:cc}
\end{align}
%of the operator $\hat{x}$ acted on
where, $\langle\hat{x}\rangle$ represents the expectation value with respect to the ket the ket $|\Psi_3\rangle$, the state after the interference. For $\theta = 0$, we get
\begin{align}
P(\theta=0) &= \frac{1}{4}\left| N_R\Phi_R + N_W\Phi_W\right|^2 \nonumber\\
&= \frac{1}{4} \left( N_R\Phi_R + N_W\Phi_W\right)\left( N_R\Phi_R^* + N_W\Phi_W^*\right) \nonumber\\
&= \frac{1}{4} \left(  \left|N_R\Phi_R\right|^2 + \left|N_W\Phi_W\right|^2 + N_RN_W|\Phi_W| \left( \Phi_R + \Phi_R^* \right) \right) \nonumber\\
&= \frac{1}{4} \left(  \left|N_R\Phi_R\right|^2 + \left|N_W\Phi_W\right|^2 + 2N_RN_W|\Phi_W| \mathrm{Re}\{\Phi_R\}  \right).
\end{align}
Here, we considered that the reference waveguide JSA varies very little over the JSA range of the resonator (Fig.~\ref{fig:qi}). However, in the presence of the phase profile in the pump, or due to the reference structure (waveguide), if there is any feature in the reference JSP, it can be included explicitly at this point to improve the accuracy of the model. In fact, flat-phase profile is not a necessary condition for our tomography. Any structure can be used as a reference if the JSP of that structure is well characterized. In our experiment, for the ease of measurement, we have used a specific length of the waveguide \cite{Faruque2019} that gives us a flat phase profile. Without any loss of generality we can assume that the $\Phi_W$ has only real part; any constant phase in $\Phi_W$ can be included in $\theta$, so we can use
\begin{align}
\Phi_W = \Phi_W^* = |\Phi_W|
\end{align}
and
\begin{align}
&\Phi_R + \Phi_R^* = 2\times\mathrm{Re}\{\Phi_R\}, \\
&\Phi_R - \Phi_R^* = 2i\times\mathrm{Im}\{\Phi_R\},
\end{align}
where, Re$\{\}$ and Im$\{\}$ indicate real and imaginary parts respectively. Similarly, we get
\begin{align}
P(\theta=\frac{\pi}{2}) &= \frac{1}{4} \left(  \left|N_R\Phi_R\right|^2 + \left|N_W\Phi_W\right|^2 - 2N_RN_W|\Phi_W| \mathrm{Re}\{\Phi_R\}  \right), \\
P(\theta=\frac{3\pi}{4}) &= \frac{1}{4} \left(  \left|N_R\Phi_R\right|^2 + \left|N_W\Phi_W\right|^2 + 2N_RN_W|\Phi_W| \mathrm{Im}\{\Phi_R\}  \right), \\
P(\theta=\frac{\pi}{4}) &= \frac{1}{4} \left(  \left|N_R\Phi_R\right|^2 + \left|N_W\Phi_W\right|^2 - 2N_RN_W|\Phi_W| \mathrm{Im}\{\Phi_R\}  \right).
\end{align}
From these equations, we get
\begin{align}
P(\theta=0) - P(\theta=\frac{\pi}{2}) = N_RN_W|\Phi_W| \mathrm{Re}\{\Phi_R\}, \\
P(\theta=\frac{3\pi}{4}) - P(\theta=\frac{\pi}{4})  = N_RN_W|\Phi_W| \mathrm{Im}\{\Phi_R\}.
\end{align}
These equations give us Eq.~\ref{eq:JSP}.\newpage

We now consider the effects of scattering losses in the resonant source and verify that even if the loss of the resonator included, the final result (Eq.~\ref{eq:JSP}) would stay the same. We model the lossy resonator following \cite{two_strategies}, introducing a phantom channel to account for the scattered photons \cite{Vernon_lossy_resonators}; in Fig \ref{fig:phantom} we sketch the system with and without the loss represented in this way. The waveguide is sufficiently short that we can continue to model it as lossless. 

Here we are dealing with a high-finesse resonator point-coupled to a channel waveguide in a weak pump regime, and our experiment relies on the measurement of coincidences. Under these conditions, the scattering loss is expected to affect only the linewidth of the JSA \cite{two_strategies, Vernon_lossy_resonators}; the above analysis should be valid even in the presence of scattering losses, provided the resonator's full linewidth is used in simulations. We verify this by comparing the derivation of Eq. \ref{eq:JSP} with and without the part of the state describing the scattered photons included. 

With the scattered photons accounted for, the state of the system before the interference is,
\begin{align}
|\Psi_2\rangle =& \int d\nu_s d\nu_i \Bigg(N_R\bigg(\Phi_R^{rr}\left(\nu_s, \nu_i \right)e^{i2\theta} c^{\dagger}_{sr}\left(\nu_s\right)c^{\dagger}_{ir}\left(\nu_i\right) + \Phi_R^{rl}\left(\nu_s, \nu_i \right)e^{i\theta}c^{\dagger}_{sr}\left(\nu_s\right) c^{\dagger}_{il}\left(\nu_i\right) \nonumber\\ &+ \Phi_R^{lr}\left(\nu_s, \nu_i \right)e^{i\theta} c^{\dagger}_{sl}\left(\nu_s\right) c^{\dagger}_{ir}\left(\nu_i\right) + \Phi_R^{ll}\left(\nu_s, \nu_i \right)c^{\dagger}_{sl}\left(\nu_s\right) c^{\dagger}_{il}\left(\nu_i\right) \bigg)  + N_W \Phi_W\left(\nu_s, \nu_i \right) d^{\dagger}_{s}\left(\nu_s\right) d^{\dagger}_{i}\left(\nu_i\right)\Bigg)|\mathrm{vac}\rangle \label{eq:JSA_parts}
\end{align}
If the scattered photons were traced out, we would recover the form of Eq. \ref{eq:no_phantom}.
We again model the beamsplitter interaction with the transformations in Eqs. \ref{eq:BSc} and \ref{eq:BSd}. In this case, only the annihilation operators associated with a real channel waveguide (labeled with the subscript $r$) will be transformed to form the interferograms,
\begin{align}
c_{kr}^{\dagger}(\nu_k) = \frac{1}{\sqrt{2}}\left( e^{\dagger}_k(\nu_k) + i f^{\dagger}_k(\nu_k) \right)
\end{align}
In Eq.~\ref{eq:JSA_parts} only the parts of the JSA that have at least one photon in the real channel will contribute to the observed interferograms; the terms in the last set of brackets will not contribute. In principle, the second and third set of brackets could contribute to the interferograms, but because we seek coincidences, these `broken pair' terms will have no effect (see Eq. \ref{eq:cc}). The only terms in \ref{eq:JSA_parts} which contribute to the interferograms are those in the first set of brackets, which has the same form as Eq. \ref{eq:no_phantom}. We can then see that the recorded interferogram will be similar to Eq.~\ref{eq:probability},
\begin{align}
P\left( \theta, \eta, \nu_s, \nu_i \right) = \frac{1}{4} \left| N_R \Phi_R^{rr}\left(\nu_s, \nu_i \right) e^{i2\theta} +  N_W \Phi_W^{rr}\left(\nu_s, \nu_i \right) \right|^2.
\end{align}

The function $\Phi_R^{rr}$ is identical to $\Phi_R$ in Eq. \ref{eq:no_phantom}, up to a constant factor \cite{two_strategies}. As mentioned in the main text, our method does not require that $N_R = N_W$, so the presence of loss should have no effect, in principle.

The above result relies on the fact that we always measure photon coincidences for our experiment. The above equation shows that the coincidence counts act as a noise filter in the weak pump regime; as long as two sources do not produce photon-pairs at the same time, the broken pairs from the resonant sources will not contribute to the noise. 

%%%%%%%%%%%%%%%%%%%%%%%%%%%%%%%%%%%%%%%%
\section{SET using Q-SpET circuit}
Following \cite{Liscidini2013}, \cite{St_JSP} and \cite{Phase_tomography}, we can reconfigure our photonic circuit for SET measurements. If the amplitude of the seed laser is $A_{seed}$ and the JSA of the stimulated emission process is $F(\nu_s,\nu_i)$, then the amplitude of the stimulated idler can be written as
\textcolor{blue}{
\begin{align}
    A_n^{SET} \propto A^*_{seed}\times F_n(\nu_s,\nu_i)
    \label{eq:Phasetomog_eqn}
\end{align}
}
where, n=\{W,R\} represents waveguide and resonator respectively. 
In this context, Eq. \ref{eq:Phasetomog_eqn} illustrates how the seed $A^*_{seed}$ imprints phase characteristics to the photon-pairs generated ($A_n^{SET}$) which would be otherwise absent in the non-stimulated process $F_n(\nu_s,\nu_i)$.
Now the arm containing the resonator source has a thermal phase element, introducing $\theta$ phase on that arm. At the final beam splitter
\begin{align}
    \mathbf{A}_{out} &= \mathrm{\mathbf{M}_{BS}}\times \mathrm{\mathbf{M}_{\theta}}\times  \mathbf{A}_{in}, \\
     \begin{bmatrix} A_{o2} \\ A_{o2} \end{bmatrix} &= \frac{e^{i\pi/2}}{\sqrt{2}} \begin{bmatrix} 1 & i \\ i & 1 \end{bmatrix} \times \begin{bmatrix} e^{i\theta} & 0 \\ 0 & 1 \end{bmatrix} \times \begin{bmatrix} A_R^{SET} \\ A_W^{SET} \end{bmatrix}.
\end{align}
Here, $\mathrm{M_{BS}}$ and $\mathrm{M_{\theta}}$ are transfer matrices of the 50:50 beam-splitter and the phase element respectively. Therefore, choosing the upper arm of the interferometer ($A_{o1}$), the interference between stimulated idlers from the waveguide and the resonator can be expresses as
\begin{align}
    I(\theta, \nu_s, \nu_i) \propto \left| N^{'}_R F_R\left(\nu_s, \nu_i \right) e^{i\theta} +  i N^{'}_W F_W\left(\nu_s, \nu_i \right) \right|^2.
\end{align}
Here, $N^{'}_n$ are constants of proportionality. This equation is very similar to Eq.~\ref{eq:cc} except the phase accumulated by the resonator arm is half that of the Eq.~\ref{eq:cc}. Therefore, following the same calculations, we have to double the values of those 4 specific values of phase to retrieve the resonator's stimulated JSA.
%%%%%%%%%%%%
\begin{figure*}[t]
\centering
\includegraphics[width=\textwidth]{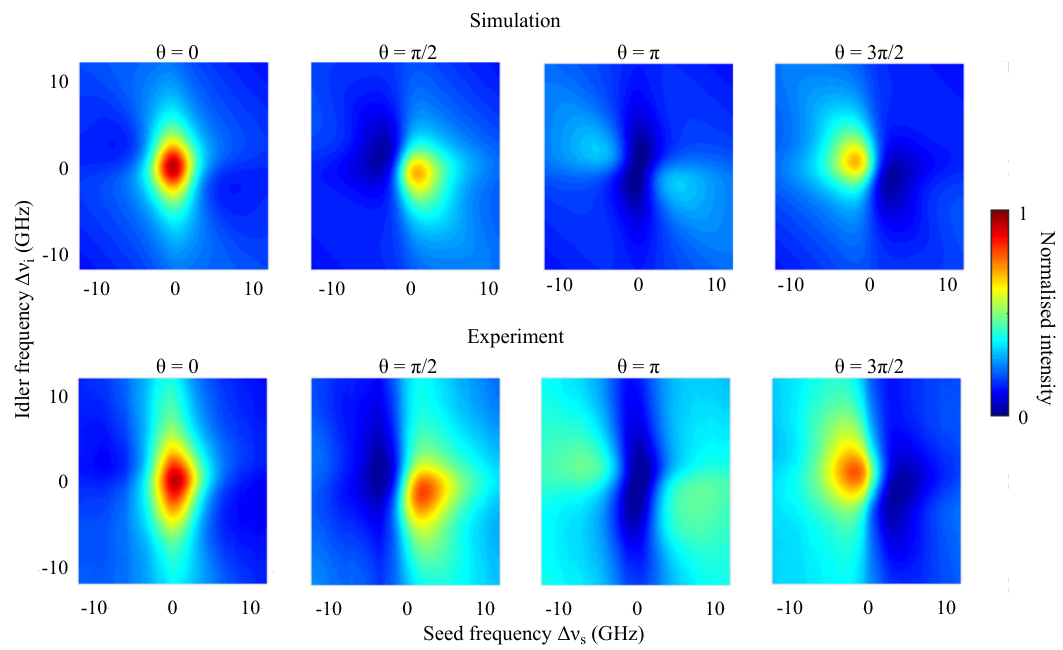}
\caption{Stimulated emission tomography (SET). Four measured interferograms at four values of $\theta = \{0,\pi/2,\pi,3\pi/2\}$, which are distinctly different from the $\theta$ values of \textcolor{red}{Q-}SpET due to the absence of the phase doubling (e.g. $2\theta$) effect of the biphoton state.
}
\label{fig:StJSP_interferograms}
\end{figure*}

\end{document}